\documentclass[fleqn,10pt]{olplainarticle}
\pdfoutput=1
\title{From far-field to near-field  micro- and nanoparticle optical trapping}

\author[1,*]{Theodoros D. Bouloumis}
\author[1,*]{S\'ile Nic Chormaic}
\affil[1]{Light-Matter Interactions for Quantum Technologies Unit, Okinawa Institute of Science and Technology Graduate University, Onna, Okinawa 904-0495, Japan}
\affil[*]{Correspondence: theodoros.bouloumis2@oist.jp; sile.nicchormaic@oist.jp}


\begin{abstract}
Optical tweezers is a very well-established technique that has developed into a standard tool for trapping and manipulating micron and submicron particles with great success in the last decades. Although the nature of light enforces restrictions on the minimum particle size that can be efficiently trapped due to Abbe's diffraction limit, scientists have managed to overcome this problem by engineering new devices that exploit near-field effects. Nowadays, metallic nanostructures can be fabricated which, under laser illumination, produce a secondary plasmonic field that does not suffer from the diffraction limit. This advance offers a great improvement in nanoparticle trapping, as it relaxes the trapping requirements compared to conventional optical tweezers. In this work, we review the fundamentals of  conventional optical tweezers, the so-called plasmonic tweezers, and related phenomena. Starting from the conception of the idea by Arthur Ashkin until recent improvements and applications, we present some of the challenges faced by these techniques as well as their future perspectives. Emphasis in this review is on the successive improvements  of the techniques and the innovative aspects that have been devised to overcome some of the main challenges.
\end{abstract}

\begin{document}

\flushbottom
\maketitle
\section{1. Introduction}
Imagine making our fingers one million times smaller and putting them "inside" the nanoworld. Now, it is easy to imagine that with these fingers we could easily grab things of similar size: dielectric nanoparticles, quantum dots with a DNA strand attached to them, proteins and viruses. More than that, we could have the ability to move them in space. Well, although we cannot modify our fingers, we have found a way to manipulate objects of that size using light!

For more than four centuries, it has been known that light can exert forces on objects \cite{RN118}. Much later, in 1873, using Maxwell's famous electromagnetic theory \cite{RN119}, the transfer of momentum from light to illuminated objects was described, resulting in the so-called radiation pressure that leads to objects moving along the direction of light propagation \cite{RN120}.   There were many experiments to follow that confirmed Poynting's calculations, but all of them were concluding with the fact that these optical forces were so small that it was difficult even to measure them, let alone utilise them in some meaningful application. However, as usually happens with science and technology, the development of techniques to allow this eventually came. Particularly, the birth of lasers around 1960 \cite{RN116,RN117} opened new possibilities and topics for research in the field of light-matter interactions.

As lasers became more and more popular in science exploration, Arthur Ashkin, in 1970, experimentally demonstrated how optical radiation forces exerted by lasers can be used to change the motion of dielectric microparticles. He even managed to trap them by creating a stable optical potential well \cite{RN2}, thus establishing the new research topic that is known today as optical tweezers. 

As always, nature follows its own rules, and soon the primary challenge for optical tweezers became apparent, i.e., the diffraction limit. It seemed to be impossible to focus light beyond the constraints imposed by this limit and, consequently, this created a restriction on the smallest size of particle that could be trapped. Subsequently, the next step was the idea to utilise surface plasmons excited on metallic nanostructures to confine light into highly intense optical fields, thus enabling superior trapping performance \cite{RN62}. The first experimental demonstration of trapping using plasmonic structures was reported by Righini et al. \cite{RN18} in 2007 and, since then, the field of plasmonic optical tweezers started developing rapidly and opened further scientific avenues for exploration. Numerous implementations arose from the research on optical forces and plasmonics which are mentioned elsewhere \cite{RN17,RN19}.

\section{2. Conventional Optical Tweezers}

The Nobel Prize in Physics 2018 was awarded (50\%) to Arthur Ashkin "\textit{for the optical tweezers and their application to biological systems}". The whole research field started when Ashkin calculated that "\textit{a power $P =1 \; W$ of cw argon laser light at $\lambda=0.5145 \; \mu m$ focussed on a lossless dielectric sphere of radius $r=\lambda$ and density $=1 \;gm/cc$ gives a radiation pressure force $F_{rad} = 2qP/c =6.6\cdot10^{-5} \; dyn$, where $q$, the fraction of light effectively reflected back, is assumed to be of order 0.1. The acceleration $=1.2 \cdot10^8 \; cm/sec^2 \cong 10^5$ times the acceleration of gravity}" \cite{RN2}. In the same work, he demonstrated the first experimental approach to test his calculations on transparent, micron-sized latex spheres in liquids and gas. Indeed, he found that the radiation pressure exerted on the particles from a focussed laser light beam was able to accelerate them along the direction of the beam and the measured velocities of the accelerated particles were in very good agreement with the theoretical predictions.

Ashkin then went even further and demonstrated trapping of particles using one laser beam and the wall of a glass cell, as well as using two counter-propagating beams with the same characteristics \cite{RN2}. However, a few years later, Ashkin et al. \cite{RN61}, reported trapping of dielectric particles (10 $\mu$m - 25 nm) using a single beam by focussing argon-laser light at 514.5 nm through a high numerical aperture objective lens ($NA=1.25$). The achievement of this is attributed to the existence of a force additional to that caused by the radiation pressure (from now on called the scattering force) which originates from the axial beam intensity gradient. It then becomes apparent that, whereas the scattering force depends on the optical intensity and has the direction of the incident beam, the gradient force depends on the intensity gradient and is directed along it from low to high intensities. This allows for optical trapping by balancing these two forces. 

The theoretical mechanism that explains this observation, depends on the relative size of the particle (radius, $r$) in respect to the wavelength of the laser light ($\lambda$). For $r \gg \lambda$, ray optics can be used and the reflection and transmission of the beam from the particle can give rise to the two forces. For $r \ll \lambda$, Rayleigh scattering is assumed and the particle is treated like a dipole in an external electromagnetic field. The two regimes are analysed below. Finally, there is the intermediate regime where the particle size is of the same order of magnitude as the wavelength. In this case, the approximations mentioned above cannot be used and, in order to evaluate the forces arising, Maxwell's stress tensor, which relates the interactions between electromagnetic forces and mechanical momentum~\cite{RN17,RN111}, should be used. To handle this complicated mathematical analysis different algorithms have been established, such as the transition matrix (T-matrix) method~\cite{RN112} and the discrete dipole approximation (DDA)~\cite{RN113}. Due to its complexity this regime is not analysed here.

\subsection{\textbf{Ray optics approximation ( \boldmath $r \gg \lambda$)}}

\begin{figure}
\center
\includegraphics[width=0.9 \textwidth]{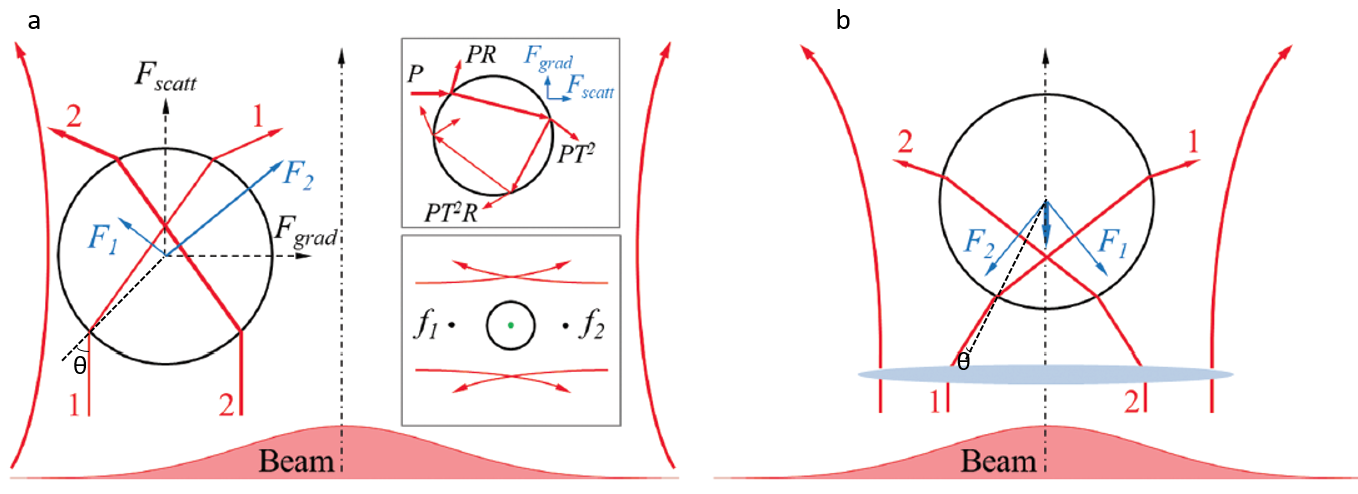} 
\caption{The scattering and gradient forces acting on a dielectric particle in the ray optics regime, arising from a free space \textbf{a} and a tightly focussed \textbf{b} Gaussian beam. In the subplot \textbf{a} the forces push the particle towards the centre of the beam and along the direction of propagation, whereas in \textbf{b} the forces drag the particle towards the focus of the beam. The upper panel in subplot \textbf{a} shows the proposed geometry for calculating the forces using Fresnel coefficients. The bottom panel shows trapping using counter-propagating beams with the same characteristics so that the scattering forces cancel each other to achieve a stable trap. Figure reproduced with permission from \cite{RN27}.}
\label{MiePrinciple}
\end{figure}

We assume spherical particles of higher refractive index than their surrounding environment, being in a liquid solution and undergoing Brownian motion. As soon as a particle, randomly moving, enters the light beam, a small fraction of light is reflected off the surface of the particle and most of it is refracted on passing through the particle (assuming no absorption). Light carries momentum and, since refraction is a light-matter interaction phenomenon, there is a momentum transfer from the photons to the particle. As is known from geometrical optics, the path of the light changes due to the refraction, resulting in a change in the momentum of the photons. Obviously, from conservation of momentum for the light-particle system, there should also be a change in the momentum of the particle and this creates a force acting on the particle, $\vec{F}=d\vec{p}/dt$. To get a first insight, we can initially assume that there is no reflection and part of the beam is refracted inside the sphere as shown in Figure \ref{MiePrinciple}a. According to the work done in \cite{RN30,RN31}, the magnitude of the force on the particle due to the momentum change of a single ray is given by

\begin{equation}
    F=\frac{n_m}{c}P,
\label{ScatF1}
\end{equation}
where $n_m$ is the refractive index of the particle, $c$ is the speed of light and $P$ is the power of the incident ray. Since the Gaussian beam has a radial intensity profile, the rays closer to the centre of the beam carry higher power (i.e., intensity); thus, the resultant force from ray 2 ($F_2$) is stronger than that from ray 1 ($F_1$), as shown in Figure \ref{MiePrinciple}a. 

The forces acting on the particle can be reduced into a longitudinal component parallel to the incident ray and a transversal one perpendicular to it. As shown in the figure, the longitudinal components of the two forces add up to create a scattering force, whereas the transversal components subtract leading to a gradient force towards the beam's higher intensity, thus moving the particle to the centre of the Gaussian beam and along its axis. Note that, for particles with a lower refractive index than the surroundings, the forces reverse and the particle moves away from the centre of the beam.

\begin{figure}
\center
\includegraphics[width=0.5 \textwidth]{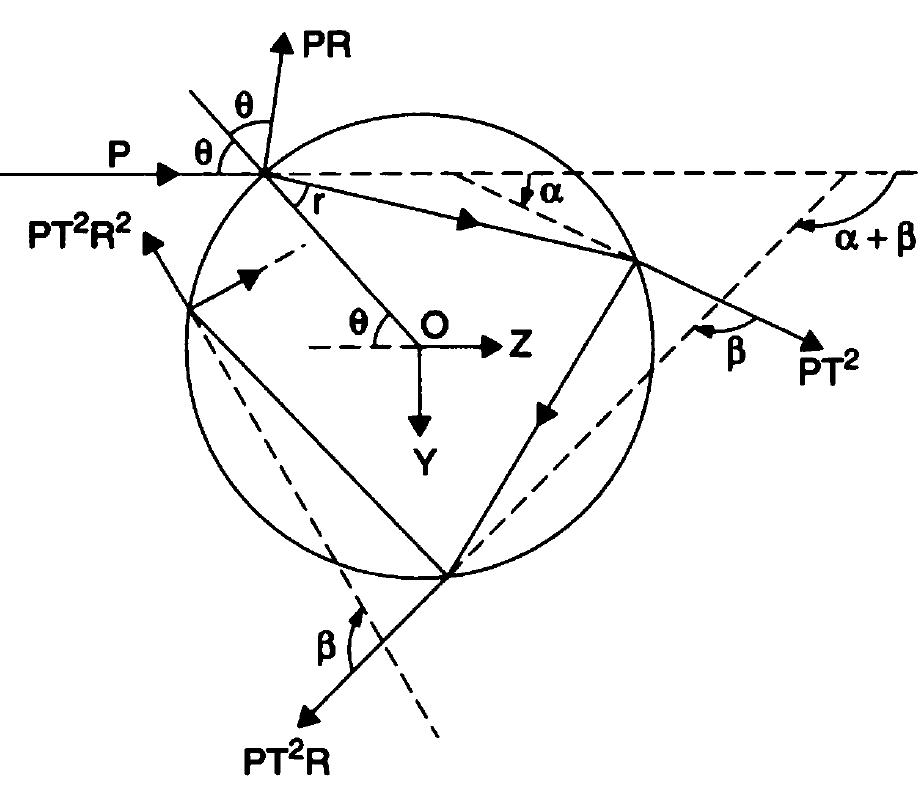} 
\caption{Calculation of the scattering and gradient forces acting on a Mie particle from a single ray, by taking into account geometrical optics and multiple reflection and transmission events, using Fresnel coefficients $R$ and $T$. Figure reproduced with permission from \cite{RN63}.}
\label{MieForceCalc}
\end{figure}

If we want to describe the process in a more accurate and mathematically rigorous way, we have to take into account multiple internal reflections and refractions of the rays, as shown in the top inset of Figure \ref{MiePrinciple}a and Figure \ref{MieForceCalc}. The forces exerted on a particle were first calculated by Roosen \cite{RN64} by considering Fresnel's reflection ($R$) and transmission ($T$) coefficients. For a detailed derivation see \cite{RN63,RN56}. The resulting forces are:

\begin{equation}
    F_{scat}=F_Z=\sum_{i=1}^{N}\dfrac{n_m \cdot P_i}{c} \left[1+R_i \cos(2\theta_i)-\dfrac{T_i^2[\cos(2\theta_i-2 r_i)+R_i \cos(2\theta_i)]}{1+R_i^2+2R_i \cos(2 r_i)} \right]
\label{OTFscat}
\end{equation}
and
\begin{equation}
    F_{grad}=F_Y=\sum_i^N \dfrac{n_m \cdot P_i}{c}\left[ R_i \sin(2 \theta_i) - \dfrac{T_i^2[\sin(2\theta_i - 2r_i)+R_i \sin(2 \theta_i)]}{1+R_i^2+2R_i \cos(2r_i)} \right],
\label{OTFgrad}
\end{equation}
where the sum is over all $N$ rays interacting with the particle, and $\theta_i$ and $r_i$ are the incidence and refraction angles, respectively, as shown in Figure \ref{MieForceCalc}. The terms in the square brackets are the dimensionless \textit{trapping efficiencies}, $Q_{scat}$ and $Q_{grad}$, and account for the efficiency of momentum transfer from the light ray to the particle. We also define the total trapping efficiency of the ray as $Q_{ray}=\sqrt{Q^2_{scat}+Q^2_{grad}}$. The Fresnel coefficients, $R$ and $T$,  depend on the polarisation of the incident rays. Therefore, the trapping efficiencies and the trapping forces will also be polarisation dependent.

\begin{figure}
\center
\includegraphics[width=0.5 \textwidth]{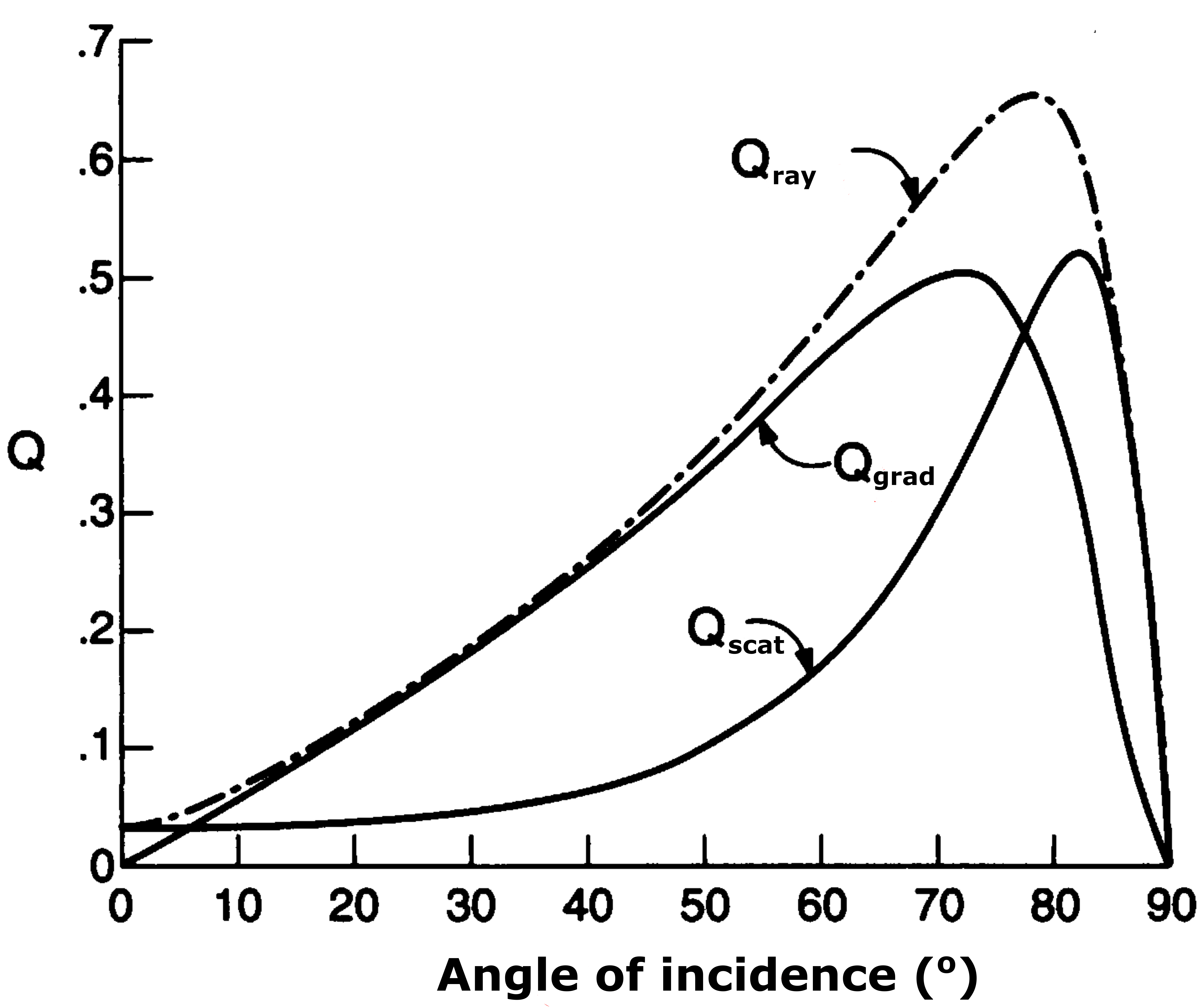} 
\caption{The calculated \textit{Q} factors for the gradient, scattering, and total trapping efficiencies for a single circularly polarised ray acting on a spherical dielectric particle of effective refractive index  $n_m=1.2$ ($n_m=n_g/n_s$), as a function of the incident angle. Figure reproduced with permission from \cite{RN63}.}
\label{Qfactors}
\end{figure}

In Figure~\ref{Qfactors}, the trapping efficiencies as a function of the ray's incident angle are plotted for circularly polarised light hitting a glass $(n_g=1.6)$ sphere in water $(n_s=1.33)$. We can see that, for incident angles smaller than $70^{\circ}$, the gradient force dominates, but as the incident angle increases, the scattering force becomes significant. This means that, for unfocussed or slightly focussed beams that have a small convergence angle, inevitably most of the rays (taking into account the Gaussian beam profile) will hit the surface of the particle with a large incident angle, $\theta$, as shown in Figure~\ref{MiePrinciple}a, thereby pushing the particle away. On the contrary, beams that are tightly focussed under a high NA objective lens, cause the rays to hit the surface of the particle with small incident angles (Figure~\ref{MiePrinciple}b). As a result, the gradient forces strongly dominate over the scattering ones and a stable trap can be established, as Ashkin et al. experimentally demonstrated~\cite{RN61}. Note that, in this case, the longitudinal component of the resulting forces always points towards the beam's focal point, as shown in Figure \ref{MiePrinciple}, leading to particle trapping close to the focal point. Ashkin's calculations \cite{RN63} confirmed that, in order to create strong, single-beam traps, high convergence angles are required. For convergence angles smaller than $\sim 30^{\circ}$, single-beam trapping is impossible. Instead, we can use two counter-propagating beams with the same characteristics, as shown in the bottom inset of Figure~\ref{MiePrinciple}a, to cancel out the scattering forces~\cite{RN2}. 

\subsection{\textbf{Dipole approximation \boldmath$(r \ll \lambda)$}}

In this case, the electric field that the particle experiences is approximately spatially constant and, assuming a dielectric particle, we can treat the entire particle as a collection of induced point dipoles in a homogeneous electric field. Early theoretical work on radiation forces and scattering effects for subwavelength dielectric media can be found in \cite{RN30,RN66}. Based on this work and the electromagnetic theory for electromagnetically induced dipoles, we can describe the optical forces and the trapping potential that arises.

The situation for a homogeneous particle can be briefly described as follows\footnote{The analysis presented is taken from \cite{RN56}.}: The oscillating electromagnetic field from the laser beam causes each of the particle's point dipoles to have a dipole moment

\begin{equation}
    \Vec{p}=\alpha_0 \Vec{E},
    \label{dipole_moment}
\end{equation}
where $\Vec{E}$ is the electric field  and $\alpha_0$ is the polarizability of the particle, given by the \textit{Clausius-Mossotti relation}

\begin{equation}
    \alpha_0=4\pi r^3 \epsilon_0 \dfrac{\epsilon_r -1}{\epsilon_r+2}.
    \label{polarizability}
\end{equation}
Here, $r$ is the particle's radius, $\epsilon_0$ is the vacuum dielectric permittivity, and $\epsilon_r$ is the particle's dielectric permittivity. The external field causes the dipoles to oscillate and, thus, radiate. Now, we have to take into account the dipole's interaction not only with the external electromagnetic field, but with its own induced scattered field as well. For that reason, the \textit{effective polarizability}, $\alpha_d$, is introduced as a radiative reaction correction to the intrinsic polarizability of the particle~\cite{RN56} and it is given by

\begin{equation}
    \alpha_d = \dfrac{\alpha_0}{1 - \frac{\epsilon_r - 1}{\epsilon_r + 2} [(k_0r)^2 + \frac{2i}{3}(k_0r)^3]},
    \label{effective_polarisability}
\end{equation}
with $k_0$ being the vacuum wavenumber.

Similarly, in order to calculate the forces acting on all the dipoles, we have to take into account the Lorentz force from the external field and the radiation forces arising from  the dipoles themselves. It is also convenient to calculate the time-averaged total force since it is the one that is observable (electromagnetic fields oscillate on the order of $\sim10^{15}$ Hz which is very fast). Rigorous calculations have been done in \cite{RN87,RN56,RN27}. According to these works, the resulting, time-averaged force acting on a dipole is given by

\begin{equation}
    \Vec{F}=\frac{1}{4}\alpha'_d \nabla |\Vec{E}|^2 + \frac{\sigma_{ext,d}}{c}\Vec{S} + \frac{c\epsilon_0 \sigma_{ext,d}}{4 \omega i} \nabla \times (\Vec{E} \times \Vec{E}^*),
    \label{dipole_force}
\end{equation}
where $\alpha'_d$ is the real part of the effective polarizability, $\omega$ is the angular frequency, $\sigma_{ext,d}$ is the \textit{extinction cross-section}, i.e., the active area of the particle that causes part of the energy of the incident electromagnetic wave to be extinguished due to scattering and absorption from the particle. It, therefore, indicates the rate of energy loss from the incident wave. $\Vec{S}$ is the time-averaged real part of the Poynting vector of the incident wave

\begin{equation}
    \Vec{S}=\frac{1}{2}Re\{\Vec{E} \times \Vec{H}^*\}.
    \label{poynting_vector}
\end{equation}
 
\paragraph{}
In Equation~(\ref{dipole_force}), we  see that the force acting on a dipole consists of three terms; the third term is called the \textit{spin-curl force} and is related to polarisation gradients in the electromagnetic field that arise when the polarisation is inhomogeneous. Using a defined polarisation of the incident beam, this term has a small value compared to the other two terms and that is why we usually neglect it in optical trapping experiments.  The second term is the scattering force pointing in the direction of the Poynting vector, $\Vec{S}$, and arises from absorption and scattering phenomena that cause momentum transfer from the field to the particle.  The first term is the gradient force and depends on the particle's polarizability and the intensity gradient of the electric field. We know that $I = \frac{1}{2}c\epsilon_0 |\vec{E}|^2$ and so the gradient force takes the form

\begin{equation}
    \vec{F}_{grad}(\vec{r}_d) = \frac{1}{2}\frac{\alpha'_d}{c\epsilon_0}\nabla I(\vec{r}_d).
    \label{gradient_force}
\end{equation}

\noindent Equation~(\ref{gradient_force}) tells us that, for particles with positive polarizability (i.e., a higher refractive index than its surrounding) this force acts towards the direction of the field's higher intensity, i.e., the focal point. At the focal point of a Gaussian beam with a beam waist, $w_0$, and radial coordinate, $\rho$, we can approximate the intensity distribution as

\begin{equation}
    I(\rho)=I_0 e^{-2\rho^2 / w_0^2},
    \label{intensity_distribution}
\end{equation}
and, for small radial displacements, we can Taylor expand to get

\begin{equation}
    I(\rho) \approx I_0 \left (1-2 \frac{\rho^2}{w_0^2} \right),
    \label{taylor_intens_distrib}
\end{equation}
and substitute into Equation~(\ref{gradient_force}). Thence,

\begin{equation}
      \vec{F}_{grad}(\vec{r}_d) = \frac{1}{2}\frac{\alpha'_d}{c\epsilon_0} \frac{\partial}{\partial\rho} \left [I_0 \left (1-2 \frac{\rho^2}{w_0^2} \right) \right] \hat{\rho} = -2\frac{\alpha'_d}{c\epsilon_0}\frac{I_0}{w_0^2}\rho \hat{\rho}.
      \label{intensity_grad_force}
\end{equation}

\noindent By comparing with the restoring force of the classical harmonic oscillator, $\vec{F}(x)=-\kappa x\hat{x}$, we get the trapping constant

\begin{equation}
    \kappa_{\rho}=2 \frac{\alpha'_d}{c\epsilon_0}\frac{I_0}{w_0^2},
    \label{trap_constant}
\end{equation}
and, by integrating Equation~(\ref{intensity_grad_force}), we get the trapping potential

\begin{equation}
    U(\rho)=\frac{1}{2} \kappa_{\rho} \rho^2,
    \label{trapping_potential}
\end{equation}
which is plotted in Figure \ref{trapping_potential}. Note that similar analysis can be used in order to obtain the potential and the trapping constant along the axial direction.

Let us now present the main limitation of conventional optical tweezers. Ashkin et al.~\cite{RN61} did some rough calculations and showed that, in order to create a stable optical trap, resisting the Brownian motion of particles in a liquid environment, a potential well as deep as $10 k_B T$ is required, where $T$ is the temperature in Kelvin. Although in some cases this is easy to achieve, for subwavelength particles, as they become smaller in size, the gradient force scales down very quickly, making it impossible to satisfy this requirement. According to Eqs.~(\ref{polarizability}) and (\ref{effective_polarisability}), $\alpha'_d \varpropto r^3$, which means that, when the radius of the particle decreases by a factor of 10, the polarizability of the particle and, consequently, the gradient force (Equation~(\ref{gradient_force})) decrease by a factor of 1,000. The trapping potential is no longer deep and tight enough to hold the particle (Figure~\ref{trapping_potential}) and the trap is inefficient. We can use Equations~\ref{polarizability}, \ref{effective_polarisability} and \ref{trap_constant} to calculate the change in the trapping stiffness if the particle has a radius of $0.8r$: 

\begin{equation}
    \Delta \kappa = \frac{\kappa_{0.8r} - \kappa_r}{\kappa_r}=\frac{\alpha'_{d0.8r} - \alpha'_{dr}}{\alpha'_{dr}}.
\end{equation}

\noindent Calculations show that for $r=100$~nm, there is a $53\%$ decrease in the trapping stiffness when the particle's radius decreases to $0.8r$. From Equation~(\ref{trap_constant}) we see that, in order to compensate for this effect and increase the trapping constant and the gradient force, we can either increase the intensity of the incident field ($I_0$) or focus tighter ($w_0$). However, even though in some cases it is experimentally possible to increase the intensity of the field by a factor of 1,000, the heat accumulation will be huge and eventually destroy the particle, especially if it is a biological sample. On the other hand, the diffraction limit allows focussing of the beam to a certain spot size and this sets a minimum on the particle size that can be successfully trapped. Additional to these limitations, as the particle becomes smaller, the viscous drag reduces and the particle undergoes more intense Brownian motion, making it easier for it to escape from the trap.
\begin{figure}[!ht]
    \center
    \includegraphics[width=0.6 \textwidth]{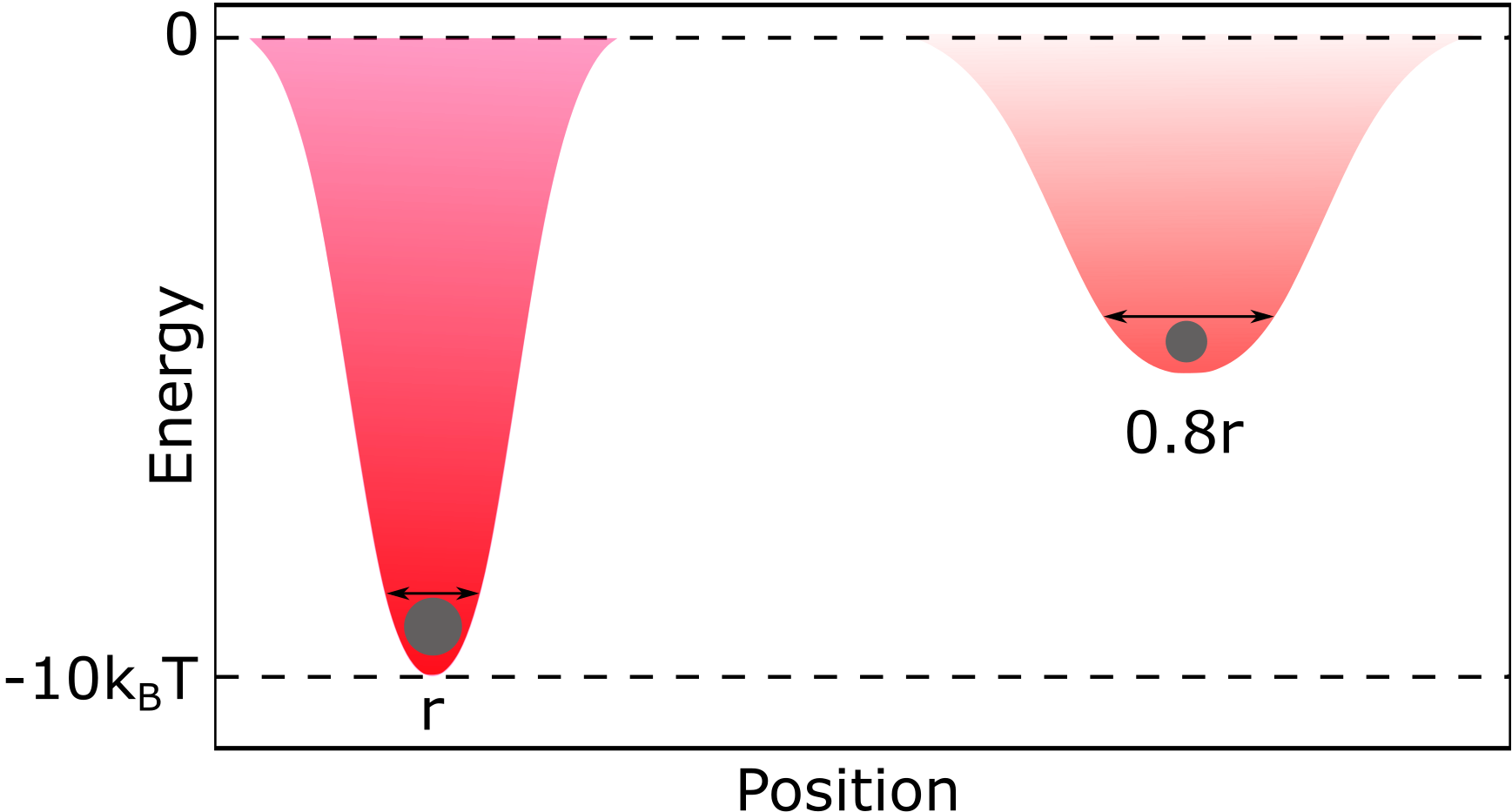}
    \caption{Graphical representation of the trapping potential wells for a polystyrene bead of radius $r$ and 0.8$r$. A small decrease in the particle's size creates a significantly shallower and broader trapping potential well, which offers much weaker confinement and, thus, the particle has a higher probability to escape from the trap.}
    \label{trapping_potential}
\end{figure}

\section{3. Plasmonic Optical Tweezers}

Recent advances in the field of optics and nano-optics have helped to overcome the diffraction limit problem by using evanescent fields instead of propagating ones; these have the intrinsic property of confinement beyond the diffraction limit. A detailed analysis can be found in~\cite{RN75,RN57}. The current trend is to use metallic nanostructures (see \cite{KNC19} for a recent review on different platforms) in which surface plasmons can be excited at resonant frequencies and that concentrate the electric field to create highly intense fields, thereby significantly increasing the trapping potential depth that a nanoparticle may experience.

\begin{figure} [!ht]
    \centering
    \includegraphics[width=0.5 \textwidth] {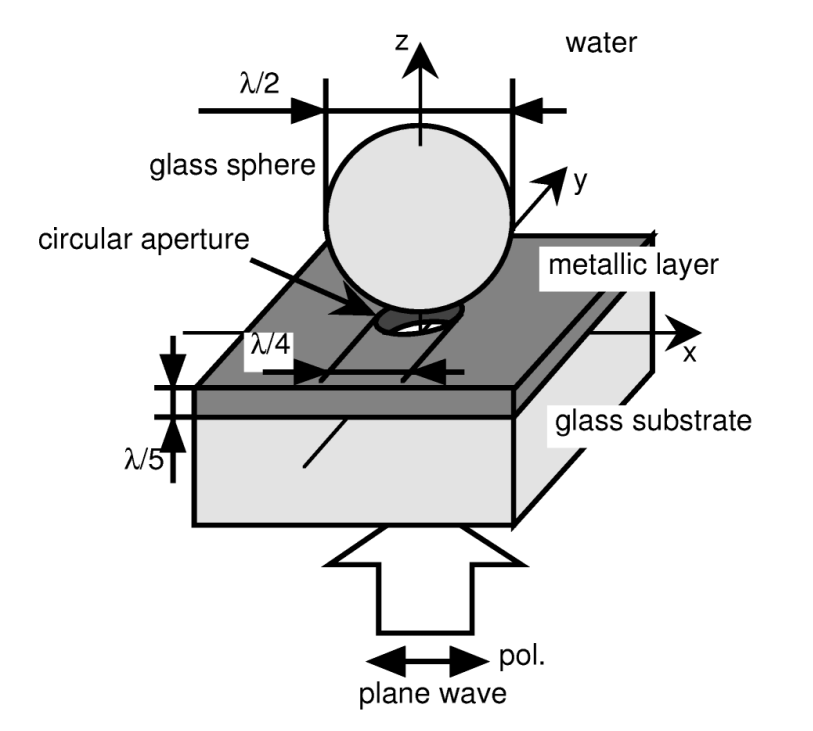}
    \caption{The geometrical model, studied by Okamoto et al., for subwavelength particle trapping, utilising the evanescent field near a metallic nanoaperture. This is the most commonly studied and reported geometry due to its simplicity to fabricate using the focussed ion beam (FIB) milling technique. Figure reproduced with permission from \cite{RN77}. }
    \label{Kawata_aperture}
\end{figure}

In 1992, Kawata et al. were the first to demonstrate \textit{"Movement of micrometer-sized particles in the evanescent field of a laser beam"}~\cite{RN76} and, in 1997, Novotny et al. were the first to theoretically propose and calculate optical trapping at the nanoscale, using enhanced evanescent fields from a laser-illuminated metallic nanotip~\cite{RN62}. Okamoto et al., around the same time, did similar work, but used a metallic nanoaperture instead of a tip~\cite{RN77}. Figure \ref{Kawata_aperture} shows their proposed geometrical model.

The advantage of using this kind of configuration comes from the fact that the incident field is no longer "responsible" for creating the trapping potential, but rather for exciting the surface plasmons (SP) on the metal/dielectric interface. The SPs, in turn, create the strong evanescent field that is "responsible" for the trapping potential. The main benefit of trapping using an evanescent field is that, by nature, it has a very high field gradient, thus exerting a large trapping force (see Equation~(\ref{gradient_force})) with no need to increase the incident intensity, thereby leading to a reduction of radiation damage to the sample. In other words, superior trapping conditions can be achieved with much lower illumination power compared to the conventional optical tweezers.

\begin{figure} [!ht]
    \centering
    \includegraphics[width = 1 \textwidth]{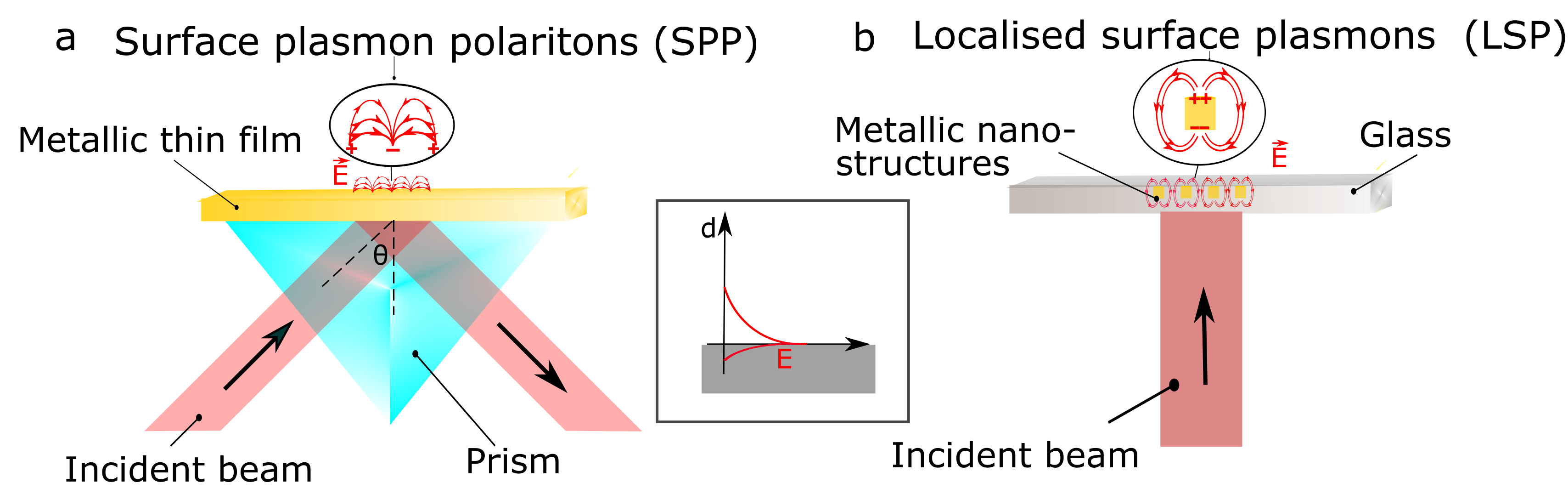}
    \caption{\textbf{a} The Kretschmann configuration, necessary to excite the surface plasmon polaritons (SPP). Light is coupled into SPPs under total internal reflection in order to compensate for the light momentum mismatch. The angle of incidence, $\theta$, controls both the scattering and the gradient force, allowing for tuning of the total trapping force. \textbf{b} Excitation of the localised surface plasmons (LSP) can happen under direct illumination of the metallic nanostuctures. The geometrical characteristics of the nanostructures on the metallic thin film define the resonance frequency and the forces that arise from the plasmonic field. Inset shows the exponential decay of the evanescent plasmonic field from the surface of the material.} 
    \label{SPP-LSP}
\end{figure}

It was then just a matter of time for the first experimentally demonstrated plasmonic optical tweezers to be reported. In 2007, Righini et al., using a geometry of total internal reflection similar to the one shown in Figure \ref{SPP-LSP}a, and a pattern consisting of 4.8 $\mu$m-diameter gold discs fabricated on glass, performed multiple trapping of 4.88 $\mu$m polystyrene colloids \cite{RN18}. Note that the laser beam was unfocussed, with a waist of about 100 $\mu$m and the intensity was more than 10 times lower than that required for conventional optical tweezers with similar characteristics. Theoretical work had been done earlier in order to study the forces arising in such a configuration~\cite{RN78}. Also, in earlier experimental work, the authors used a photonic force microscope to measure the plasmon radiation forces acting on polystyrene beads at the \textit{localised surface plasmon} resonance. They reported forces 40 times stronger than those obtained in the absence of SP excitation \cite{RN21}.

Now, it is important to mention the two distinct types of surface plasmons. Surface plasmon polaritons (SPP) are propagating electromagnetic surface waves that appear at the metal/dielectric interface due to the motion of the metal's free electrons driven by the incident electromagnetic field. They are evanescent modes and, thus, they produce localised fields with a high intensity decaying exponentially away from the metal surface. Due to the very large intensity gradient, they exert strong gradient forces on the trapped particles (see Equation~(\ref{gradient_force})), thereby producing stable traps. However, because SPPs are pure evanescent modes, direct coupling to propagating light is not possible and, in order to excite them, a different geometrical approach is required. The most common experimental method is the Kretschmann configuration, which is shown in Figure~\ref{SPP-LSP}a. Light is coupled into SPPs under total internal reflection in order to compensate for the light momentum mismatch. The crucial parameter in this configuration is the angle of incidence, $\theta$, which controls both the scattering and the gradient force, allowing for tuning of the total trapping force.

In contrast, localised surface plasmons (LSP) are related to the bound electrons that are present near to nanoapertures or nanoparticles much smaller than the wavelength of the electromagnetic field. Bound electrons are susceptible to a damping oscillation due to the nucleus attraction and, as a result, they have a characteristic resonance frequency, unlike  SPPs that can be excited over a wide range of frequencies. The benefits of LSPs are that they can directly couple to propagating light and their resonance frequency can be tuned by changing the size and the shape of the nanoaperture/nanoparticle (Figure~\ref{SPP-LSP} b). In a theoretical work done on LSPs, the dramatic dependence of the strength of the excited evanescent field on the frequency of the incident electromagnetic field was presented \cite{RN78}. Detailed mathematical analysis on the excitation of surface plasmons and the forces arising can be found in Ref. \cite{RN56} and \cite{Maier}.

To date, many different configurations have been reported using SPs for efficient trapping of subwavelength particles, such as plasmonic \textit{nanodots} \cite{RN81}, \textit{nano-antennas} \cite{RN82}, \textit{nanocavities} \cite{RN15} and \textit{nano-apertures} of different shapes and sizes \cite{RN83,RN84}, reporting very low incident power. Note that, in these cases, rigorous calculations need to be done beforehand, in order to determine crucial parameters such as resonance wavelength and polarisation in compliance with the plasmonic field excitation requirements. Additionally, the fabrication of those nanostructures can also be a challenging task. The most popular techniques to fabricate structures of these sizes is focussed ion beam (FIB) milling and  electron beam lithography (EBL), with a resolution of about 10 nm.

To conclude, we emphasize that the principle of trapping using surface plasmons is the same as in the case of conventional optical tweezers, in the sense that again we have to find the appropriate balance between the gradient and the scattering force in order to achieve a stable trap. The improvement comes from the fact that, in plasmonic optical tweezers, the excited plasmonic field offers stronger gradient forces and a better control over them, as explained above. However, due to the conductive nature of metals, the excitation of SPs is connected with heat induction and dissipation to the surrounding environment and these can increase the destabilisation forces.

\subsection{\textbf{Self-Induced Back Action Effect}}

\begin{figure} [!ht]
    \centering
    \includegraphics[width = 0.8 \textwidth]{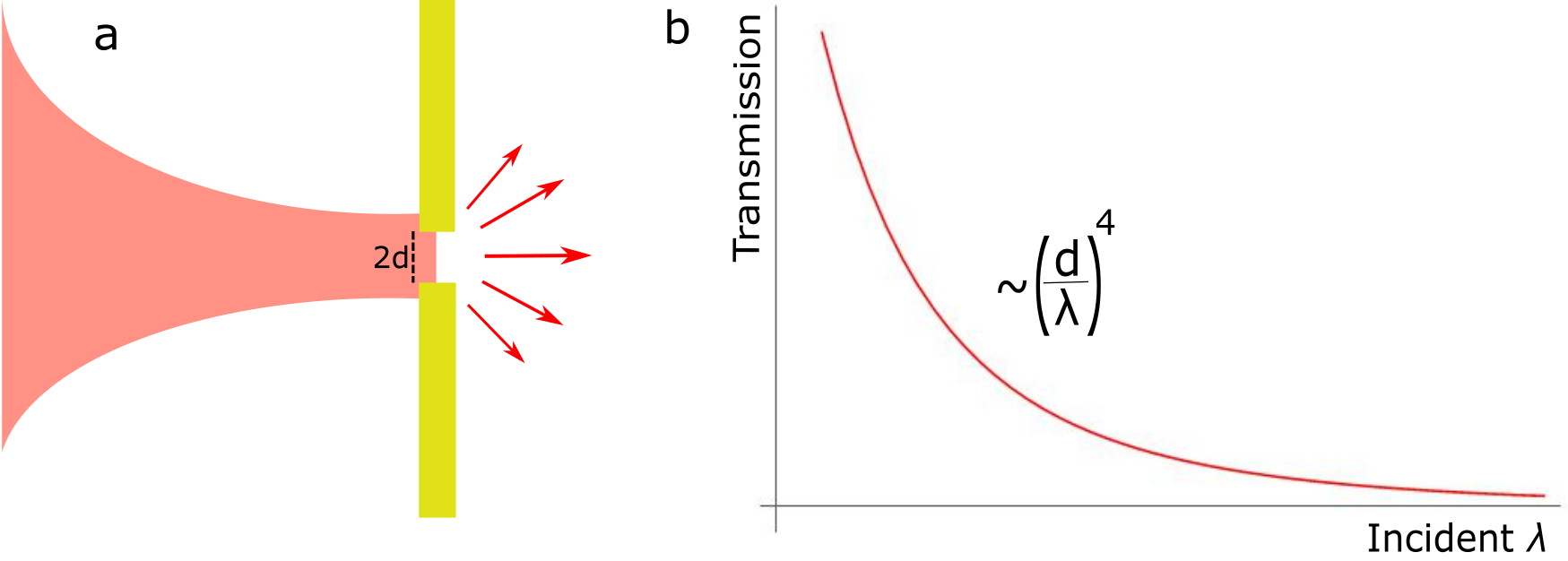}
  \caption{\textbf{a} Diffraction and transmission of visible light through a subwavelength circular hole of radius, $d$, on a perfectly conducting and infinitely thin metallic film. \textbf{b} According to Bethe's calculation the transmission of incident light $\lambda$ scales as 
  $(d/ \lambda)^4$. Figure inspired by \cite{RN6}.}
    \label{Bethe_theory}
\end{figure}

\begin{figure} [!ht]
    \centering
    \includegraphics[width = 1 \textwidth]{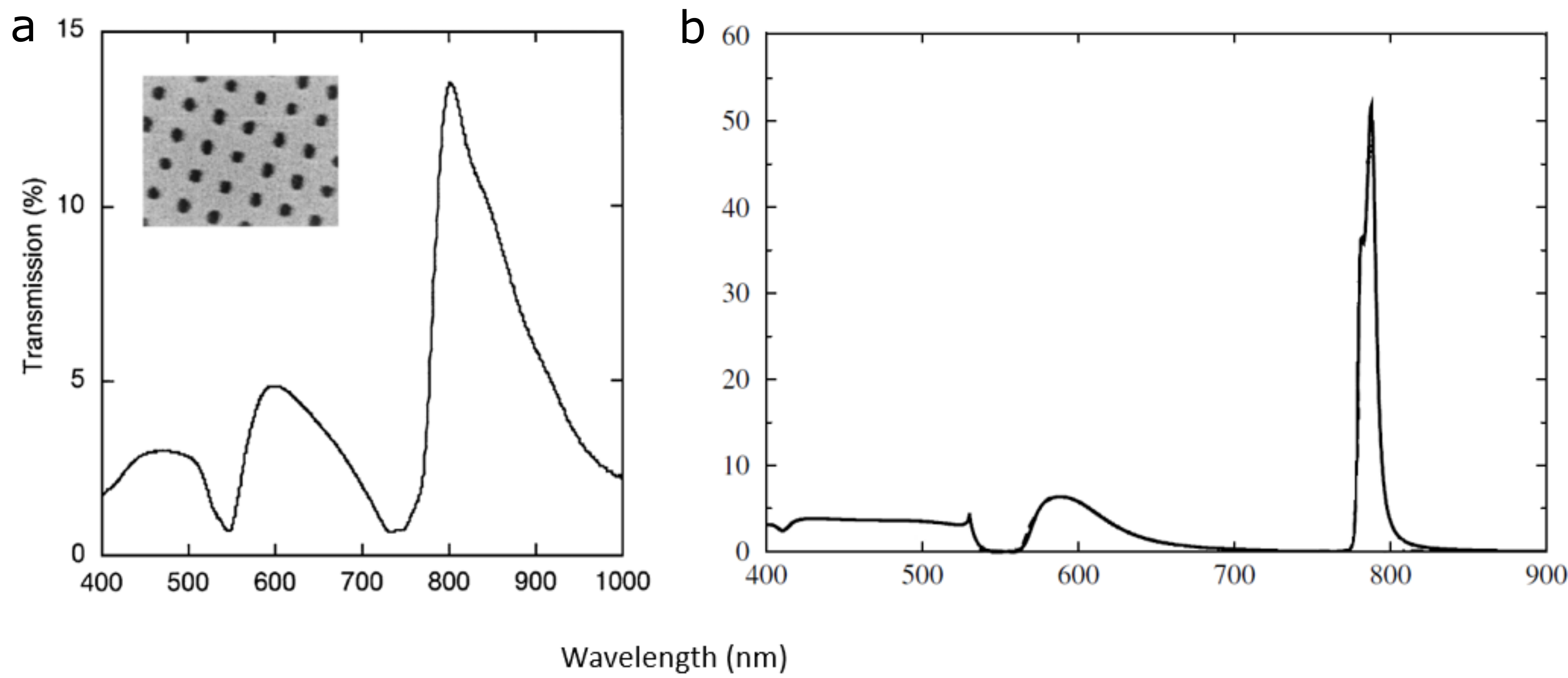}
    \caption{Experimental (left) and theoretical (right) plots of the normalised transmittance as a function of the incident light wavelength, for a square array of holes on a Ag thin film with thickness $h=$320~nm. The diameter of the holes was 280 nm and the lattice constant 750 nm. Inset in the left plot shows an SEM image of the nano-hole lattice. Image reproduced with permission from \cite{RN90}.}
    \label{transm_peaks}
\end{figure}

The diffraction and transmission of light through a single, subwavelength-sized circular hole on a metallic surface was first theoretically studied in 1944 by Bethe~\cite{RN5}. Assuming a perfectly conducting and infinitely thin material, Bethe calculated that the transmitted light would scale as $T \varpropto \left ( d / \lambda \right ) ^4$, where $d$ is the radius of the hole and $\lambda$ the wavelength of the incident light, as illustrated in Figure~\ref{Bethe_theory}. In 1998, the remarkable phenomenon of the \textit{extraordinary transmission of light} was experimentally observed by Ebbesen et al. \cite{RN88}, when they studied the effects of the geometrical characteristics of an array of tiny holes, drilled on different metallic films, on the transmission of light. In that work, UV-Vis-NIR spectrophotometry of the array was performed and it revealed the existence of maxima in the transmission intensity, see Figure~\ref{transm_peaks}a, with much higher values than predicted from Bethe's theory. These maxima could not be explained simply by diffraction theory and they were associated with the resonant frequencies of the excited surface plasmons of the metal. A detailed theoretical explanation of this observation was provided a few years later \cite{RN90, RN85}, where the calculated transmission maxima were in very good agreement with the experiment (Figure \ref{transm_peaks}b).

In addition, in his theoretical work, Abajo investigated the case of filling the nano-hole with a dielectric material of high refractive index (Si). He found out that the transmission cross-section at the resonance frequency was almost three times higher than without the filling, leading to increased transmission \cite{RN85}. The increase of the refractive index at the nano-hole causes the wavelength of the light to decrease:

\begin{equation}
    n_{fill}=\frac{c}{c_{fill}} > 1 \Leftrightarrow n_{fill}=\frac{\lambda}{\lambda_{fill}} > 1 \Leftrightarrow \lambda_{fill} < \lambda.
\end{equation}

\begin{figure}
    \centering
    \includegraphics[width = 0.8 \textwidth]{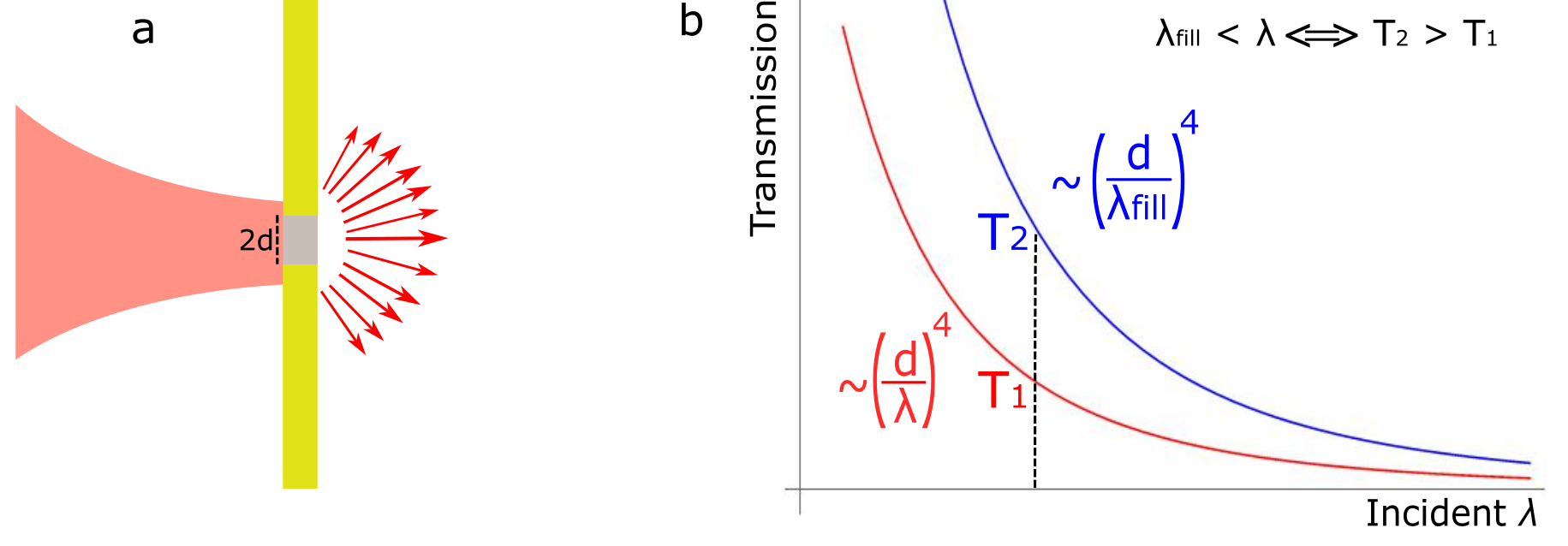}
    \caption{\textbf{a} The presence of a dielectric material of higher refractive index than the surroundings causes an increase in the transmission of light through the subwavelength aperture. \textbf{b} Shift of the transmission line due to existence of the dielectric material, leading to higher transmission. Figure inspired by \cite{RN6}.}
    \label{Bethe_extratransmission}
\end{figure}

\noindent A consequence of this effect is that, according to Bethe's theory, the transmission of light through the subwavelength aperture increases, a phenomenon also known as dielectric loading. Figure~\ref{Bethe_extratransmission} shows how the wavelength shift causes a significant increase in the transmission of light.

As mentioned in the previous section, at around the same time, the use of plasmonics for enhanced optical trapping had started to attract attention and, despite how promising they might seem for trapping subwavelength particles, it became apparent that their use is limited to particles with a minimum diameter around 100 nm due to photothermal effects \cite{RN22}.

It was the combination of these two different studies on plasmonics that brought the realisation that the resonance frequency of the excited plasmonic field is very sensitive to changes of the local refractive index. Thus, by proper engineering of the plasmonic structure, the trapped particle itself could actively contribute to its own trapping potential in a dynamical way \cite{RN91}. This plasmonic structure - particle interaction promised high tunability of the trapping potential, which was no longer a static. This gave rise to the \textit{self-induced back action} (SIBA) effect and the first experimental trapping utilising this effect, where polystyrene spheres of 100 and 50 nm size were successfully trapped with incident powers as low as 0.7 and 1.9 mW, respectively \cite{RN7}, pushing further the boundaries of plasmonic nanotweezers.

A comprehensive mathematical analysis of the SIBA effect has been done by Neu\-meier et al.~\cite{RN11}, for a small dielectric particle trapped in a plasmonic nanocavity. They demonstrated the additional restoring forces that act on the particle as it tries to escape from the trap. Below, we present the basic principle of the SIBA effect following the analysis done in~\cite{RN22}.

The gradient force experienced by a nanosphere with radius $r \ll \lambda$ of the incident light, is given by Equation~(\ref{gradient_force}), as mentioned previously. If we assume small displacements of the particle from the centre of the trap ($|x|\ll \lambda$), then its equation of motion inside the trap is given by

\begin{equation}
    \gamma \dot{x}(t) + \kappa_{tot} x(t) = \xi(t),
    \label{overdamped_Langevin}
\end{equation}
where $\gamma$ is the viscous damping \cite{RN121}, assuming that the particle exists in a liquid environment, $\kappa_{tot}$ is the stiffness of the trap, indicating how strongly the particle is confined in the trap, and $\xi$ represents thermal fluctuations~\cite{RN122}.  Due to the coupling between the cavity and the particle, the latter causes the plasmon resonance frequency of the cavity to shift by $\delta \omega_0 (x_p)$, where $x_p$ indicates the frequency dependence on the particle's position. Then, for a cavity with mode volume, $V_m$, and intensity profile, $f(x_p)$, normalised to 1 for maximum intensity, the perturbation theory for shifts much smaller than the cavity eigenfrequency, $\omega_c$, yields 

\begin{equation}
    \delta \omega_0 (x_p) = \omega_c \frac{\alpha_d}{2V_m \epsilon_0} f(x_p),
    \label{freq_shift}
\end{equation}
with $\alpha_d$ being the effective polarizability, from Equation~(\ref{effective_polarisability}). Note that the magnitude of the shift strongly depends on the relative size of the particle and the cavity and, as expected from Equation~(\ref{freq_shift}), a decrease in the particle's size (decrease in $\alpha_d$, see Eqs.~(\ref{polarizability}), (\ref{effective_polarisability})), decreases the magnitude of the shift~\cite{RN12}. 

Now, for incident laser frequency, $\omega$, and $\Delta \equiv \omega - \omega_c$ being the cavity detuning, the intracavity intensity, $I(\omega)$, is given, on Taylor expansion, as

\begin{equation}
    I(\omega)= I_0 \frac{(\Gamma/2)^2}{(\Delta - \delta \omega_0)^2 + (\Gamma/2)^2)} \approx I_{opt} - \frac{2 \delta \omega_0 (x_p) \Delta}{\Delta^2 + (\Gamma/2)^2}I_{opt} + ... 
    \label{intracavity_intensity}
\end{equation}
where $\Gamma$ is the cavity linewidth and $I_{opt}=I_0 \frac{(\Gamma/2)^2}{\Delta^2 + (\Gamma/2)^2}$ is the empty cavity profile.

As can been seen from Equation~(\ref{intracavity_intensity}), the intensity of light inside the cavity, to a first order approximation, is given by the term related to the intensity of the empty cavity, plus the one related to the frequency shift caused by the presence of the particle. This second term is the one that causes the SIBA effect and modifies the optical potential. Following Equation~(\ref{intracavity_intensity}), we can also write the total trapping stiffness, $\kappa_{tot}$, as 

\begin{equation}
    \kappa_{tot}= \kappa_{opt} + \kappa_{SIBA},
    \label{trap_stif}
\end{equation}
where $\kappa_{opt}$ is constant and depends on the cavity resonance profile and $\kappa_{SIBA}$ is a function of the particle's displacement. According to Neumeir et al., in order to optimise $\kappa_{SIBA}$, the cavity has to be constructed such that the \textit{back-action parameter}, $\upsilon = \delta \omega_0 (x_p) / \Gamma$, is maximised~\cite{RN11}. This means that, while the particle is trapped in the centre of the trap, the resonance shift is such that the photon flux from the cavity is less than the maximum possible. As a consequence, when the particle moves away from the centre of the trap, the resonance shift causes the photon flux to increase and, thus, the intensity of the transmitted light increases. From Equation~(\ref{gradient_force}) an increase to the intensity leads to an increase to the gradient force, which restores the particle back to the centre of the trap. Then, the photon flux and the intensity decrease and, again, the particle tends to move away from the trap centre. This kind of feedback is referred to as \textit{"optomechanical coupling"} because there is a continuous response between light and mechanical motion. The field of optomechanics in plasmonics is rather unexplored and, to our knowledge, there is only one work that reports an optomechanical coupling constant~\cite{RN22}. This optomechanical coupling not only relaxes the requirements for high power trapping, but also prevents the sample from overheating since most of the time the particle is trapped using a low intensity. It remains open to exploration to find ways to increase the optomechanical coupling constant and to achieve even higher particle confinement and motion transduction.

\begin{figure} [!ht]
    \centering
    \includegraphics[width = 0.8 \textwidth]{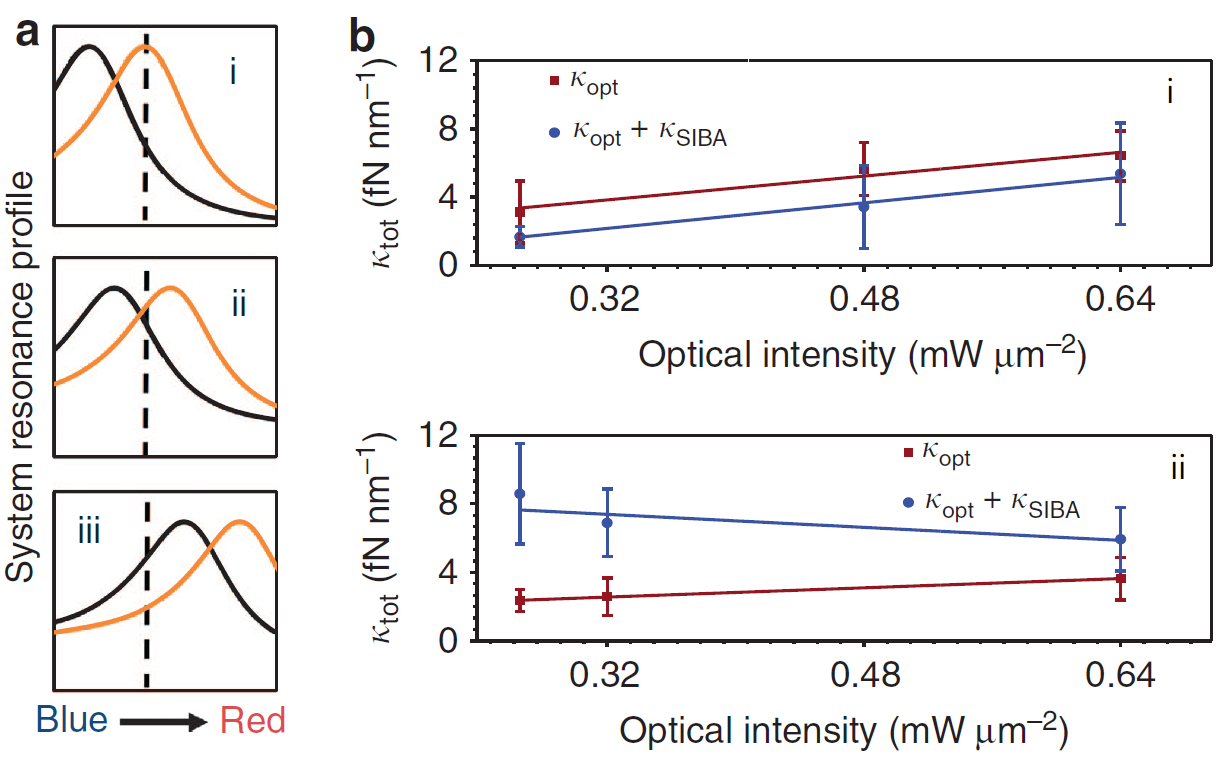}
    \caption{ \textbf{a}. Three different cases of the resonance profile of a plasmonic cavity (black line). In case (i), (ii), and (iii) the cavity is designed to be blue-detuned, slightly blue-detuned and red-detuned, respectively, from the excitation laser (vertical dashed line). The orange line represents the frequency shift of the resonance of the cavity due to the existence of the particle near by it. In case (ii), the SIBA effect has a positive influence on the trap and, whenever the particle tries to escape, the photon flux (intensity) increases, which causes, increase to the gradient force. \textbf{b}. Experimental measurements of the total trapping stiffness as a function of the incident laser intensity. \textbf{b}(i) corresponds to case \textbf{a}(i), where as the intensity decreases both the total and the empty cavity trapping stiffness decrease, making the trap inefficient. \textbf{b}(ii) corresponds to case \textbf{a}(ii), and clearly shows that, as the laser intensity decreases, the empty cavity trapping stiffness decreases, but the total stiffness increases due to the positive contribution from the SIBA effect, thus achieving a stable, self-adjustable trap. Figures taken from \cite{RN22}, under Creative Commons license: CC BY-NC-ND 4.0 https://creativecommons.org/licenses/by-nc-nd/4.0/.} 
    \label{SIBA_resonance}
\end{figure}

In Figure~\ref{SIBA_resonance}a, the vertical dashed line represents the excitation laser wavelength, the black lineshape is the empty cavity mode resonance, and the orange is the shifted one due to particle trapping. In the first case (Figure~\ref{SIBA_resonance}a(i), blue-shifted) the cavity resonance is set to be blue-detuned  from the excitation laser, such that when the particle is trapped, the resonance red-shift increases the photon flux to the maximum value and the gradient force reaches a maximum. However, when the particle moves away from the centre of the trap, the lineshape blue shifts towards the empty cavity resonance and the intensity decreases. In order to increase the gradient force, the power of the laser has to, externally, be increased, thus, it is not the most efficient scenario for trapping. In the case where the empty mode resonance is slightly blue-detuned (Figure~\ref{SIBA_resonance}a(ii), resonance), the red-shifting due to a trapping event creates a symmetrical lineshape around the laser wavelength. As the particle tries to escape from the trap, the resonance moves towards the empty mode value and causes an increase in the photon flux and the light intensity of light, thus increasing the gradient force. Finally, in the case where the empty cavity resonance is designed to be red-detuned from the excitation laser (Figure~\ref{SIBA_resonance}a(iii), red-shifted), the trapped object further red-shifts the resonance, leading to significant reduction in the intensity and the gradient force. In this configuration, the trapping becomes very inefficient and an increase to the laser power is necessary to keep the particle in the trap.

The intensity required to keep the particle efficiently in the trap is less in the second case where the SIBA effect contributes to an increase in the total trapping stiffness. This was also experimentally observed for the first time by Mestres et al. \cite{RN22}. Figure \ref{SIBA_resonance}b shows the experimental data that confirm the superior trapping efficiency of a plasmonic cavity, designed to be slightly blue-detuned from the excitation wavelength (Figure~\ref{SIBA_resonance}a(ii), b(ii)).   The remarkable effect of SIBA is now apparent and, by proper design of the plasmonic structure, we can have a larger trapping stiffness at a lower laser power, thereby reducing heat transfer to the specimen. 

\section{4. Conclusions and Future Perspectives}

There are many open questions and challenges to overcome in order to optimise the plasmonic tweezers and make them able to efficiently trap  particles in the range of less than 10 nm and biological samples such as proteins, viruses and DNA. In our opinion, one of these important questions is whether it is possible to enhance the SIBA effect itself so that we achieve even better optomechanical response of the system and superior trapping performance. 

It is amazing to see how a simple idea can be transformed into a powerful technique for controlling matter, finally leading to the awarding of a  Nobel Prize. But it is even more amazing to see people devoting themselves to solving a particular problem and pushing the boundaries of science into the unknown to make it possible. Optical tweezers-based platforms are today, thanks to all these scientistific efforts, widely used in the fields of physics~\cite{RN123}, biomedicine~\cite{RN124,RN125,RN126}, chemistry~\cite{RN127} and many more. In many cases, it serves as a tool to manipulate matter while doing other measurements, for example Raman spectroscopy of biological samples~\cite{RN128, RN129, RN130}. This gives us a great advantage towards the exploration of the nanoworld and the advancement of the nanotechnologies. What is next?

\paragraph{\textbf{Acknowledgements}}
The authors would like to thank Okinawa Institute of Science and Technology Graduate University for funding, and D. G. Kotsifaki, P. Reece, P. Okada, F. Pauly, and G. Tkachenko for constructive comments.


\bibliography{biblio}

\begin{thebibliography}{55}
\providecommand{\natexlab}[1]{#1}
\providecommand{\url}[1]{\texttt{#1}}
\expandafter\ifx\csname urlstyle\endcsname\relax
  \providecommand{\doi}[1]{doi: #1}\else
  \providecommand{\doi}{doi: \begingroup \urlstyle{rm}\Url}\fi

\bibitem[Kepler(1619)]{RN118}
J.~Kepler.
\newblock \emph{De cometis libelli tres}.
\newblock Typis Andreæ Apergeri, sumptibus Sebastiani Mylii bibliopolæ
  Augustani, 1619.
\newblock URL \url{https://books.google.co.jp/books?id=5Z1BQAAACAAJ}.

\bibitem[Maxwell(1873)]{RN119}
J.C. Maxwell.
\newblock \emph{A Treatise on Electricity and Magnetism}, volume~2.
\newblock Oxford, 1873.

\bibitem[Poynting and William(1884)]{RN120}
J.~H. Poynting and S.~J.XV William.
\newblock On the transfer of energy in the electromagnetic field.
\newblock \emph{Phil. Trans. R. Soc.}, 175, 1884.

\bibitem[Maiman(1960)]{RN116}
T.~H. Maiman.
\newblock Stimulated optical radiation in ruby.
\newblock \emph{Nature}, 187\penalty0 (4736):\penalty0 493--494, 1960.
\newblock ISSN 1476-4687.
\newblock \doi{10.1038/187493a0}.
\newblock URL \url{https://doi.org/10.1038/187493a0}.

\bibitem[Schawlow and Townes(1958)]{RN117}
A.~L. Schawlow and C.~H. Townes.
\newblock Infrared and optical masers.
\newblock \emph{Physical Review}, 112\penalty0 (6):\penalty0 1940--1949, 1958.
\newblock ISSN 0031-899X.
\newblock \doi{10.1103/PhysRev.112.1940}.

\bibitem[Ashkin(1970)]{RN2}
A.~Ashkin.
\newblock Acceleration and trapping of particles by radiation pessure.
\newblock \emph{Phys. Rev. Lett.}, 24\penalty0 (4):\penalty0 156--159, 1970.
\newblock ISSN 0031-9007.
\newblock \doi{10.1103/PhysRevLett.24.156}.

\bibitem[Novotny et~al.(1997)Novotny, Bian, and Xie]{RN62}
Lukas Novotny, Randy~X. Bian, and X.~Sunney Xie.
\newblock Theory of nanometric optical tweezers.
\newblock \emph{Phys. Rev. Lett.}, 79\penalty0 (4):\penalty0 645--648, 1997.
\newblock \doi{10.1103/PhysRevLett.79.645}.
\newblock URL \url{https://link.aps.org/doi/10.1103/PhysRevLett.79.645}.

\bibitem[Righini et~al.(2007)Righini, Zelenina, Girard, and Quidant]{RN18}
Maurizio Righini, Anna~S. Zelenina, Christian Girard, and Romain Quidant.
\newblock Parallel and selective trapping in a patterned plasmonic landscape.
\newblock \emph{Nat. Photonics}, 3\penalty0 (7):\penalty0 477--480, 2007.
\newblock ISSN 1745-2473 1745-2481.
\newblock \doi{10.1038/nphys624}.

\bibitem[Marago et~al.(2013)Marago, Jones, Gucciardi, Volpe, and Ferrari]{RN17}
O.~M. Marago, P.~H. Jones, P.~G. Gucciardi, G.~Volpe, and A.~C. Ferrari.
\newblock Optical trapping and manipulation of nanostructures.
\newblock \emph{Nat. Nanotechnol.}, 8\penalty0 (11):\penalty0 807--19, 2013.
\newblock ISSN 1748-3395 (Electronic) 1748-3387 (Linking).
\newblock \doi{10.1038/nnano.2013.208}.
\newblock URL \url{https://www.ncbi.nlm.nih.gov/pubmed/24202536}.

\bibitem[Juan et~al.(2011)Juan, Righini, and Quidant]{RN19}
Mathieu~L. Juan, Maurizio Righini, and Romain Quidant.
\newblock Plasmon nano-optical tweezers.
\newblock \emph{Nature Photon.}, 5\penalty0 (6):\penalty0 349--356, 2011.
\newblock ISSN 1749-4885 1749-4893.
\newblock \doi{10.1038/nphoton.2011.56}.

\bibitem[Ashkin et~al.(1986)Ashkin, Dziedzic, Bjorkholm, and Chu]{RN61}
A.~Ashkin, J.~M. Dziedzic, J.~E. Bjorkholm, and Steven Chu.
\newblock Observation of a single-beam gradient force optical trap for
  dielectric particles.
\newblock \emph{Opt. Lett.}, 11\penalty0 (5):\penalty0 288--290, 1986.
\newblock \doi{10.1364/OL.11.000288}.
\newblock URL \url{http://ol.osa.org/abstract.cfm?URI=ol-11-5-288}.

\bibitem[Pfeifer et~al.(2007)Pfeifer, Nieminen, Heckenberg, and
  Rubinsztein-Dunlop]{RN111}
Robert N.~C. Pfeifer, Timo~A. Nieminen, Norman~R. Heckenberg, and Halina
  Rubinsztein-Dunlop.
\newblock Colloquium: Momentum of an electromagnetic wave in dielectric media.
\newblock \emph{Rev. Mod. Phys.}, 79:\penalty0 1197--1216, Oct 2007.
\newblock \doi{10.1103/RevModPhys.79.1197}.
\newblock URL \url{https://link.aps.org/doi/10.1103/RevModPhys.79.1197}.

\bibitem[Borghese et~al.(2007)Borghese, Denti, and Saija]{RN112}
Ferdinando Borghese, Paolo Denti, and Rosalba Saija.
\newblock \emph{Scattering from model nonspherical particles}.
\newblock Springer-Verlag Berlin Heidelberg, 2007.
\newblock \doi{10.1007/978-3-540-37414-5}.

\bibitem[Draine and Flatau(1994)]{RN113}
Bruce~T. Draine and Piotr~J. Flatau.
\newblock Discrete-dipole approximation for scattering calculations.
\newblock \emph{Journal of the Optical Society of America A}, 11\penalty0
  (4):\penalty0 1491--1499, 1994.
\newblock \doi{10.1364/JOSAA.11.001491}.
\newblock URL \url{http://josaa.osa.org/abstract.cfm?URI=josaa-11-4-1491}.

\bibitem[Gordon(1973)]{RN30}
James~P. Gordon.
\newblock Radiation forces and momenta in dielectric media.
\newblock \emph{Phys. Rev. A}, 8\penalty0 (1):\penalty0 14--21, 1973.
\newblock ISSN 0556-2791.
\newblock \doi{10.1103/PhysRevA.8.14}.

\bibitem[Ashkin and Dziedzic(1973)]{RN31}
A.~Ashkin and J.~M. Dziedzic.
\newblock Radiation pressure on a free liquid surface.
\newblock \emph{Phys. Rev. Lett.}, 30\penalty0 (4):\penalty0 139--142, 1973.
\newblock ISSN 0031-9007.
\newblock \doi{10.1103/PhysRevLett.30.139}.

\bibitem[Bradac(2018)]{RN27}
Carlo Bradac.
\newblock Nanoscale optical trapping: A review.
\newblock \emph{Adv. Opt. Mater.}, 6\penalty0 (12), 2018.
\newblock ISSN 21951071.
\newblock \doi{10.1002/adom.201800005}.

\bibitem[Roosen(1979)]{RN64}
Gérald Roosen.
\newblock La lévitation optique de sphères.
\newblock \emph{Can. J. Phys.}, 57\penalty0 (9):\penalty0 1260--1279, 1979.
\newblock ISSN 0008-4204.
\newblock \doi{10.1139/p79-175}.
\newblock URL \url{https://doi.org/10.1139/p79-175}.

\bibitem[Ashkin(1992)]{RN63}
A.~Ashkin.
\newblock Forces of a single-beam gradient laser trap on a dielectric sphere in
  the ray optics regime.
\newblock \emph{Biophys. J.}, 61\penalty0 (2):\penalty0 569--582, 1992.
\newblock ISSN 0006-3495.
\newblock \doi{https://doi.org/10.1016/S0006-3495(92)81860-X}.
\newblock URL
  \url{http://www.sciencedirect.com/science/article/pii/S000634959281860X}.

\bibitem[Jones et~al.(2015)Jones, Marago, and Volpe]{RN56}
Philip Jones, Onofrio~M. Marago, and Giovanni Volpe.
\newblock \emph{Optical Tweezers: Principles \& Applications}.
\newblock Cambridge University Press, Cambridge, 2015.
\newblock ISBN 9781107051164.

\bibitem[M.~Purcell and R.~Pennypacker(1973)]{RN66}
Edward M.~Purcell and Carlton R.~Pennypacker.
\newblock Scattering and absorption of light by nonspherical dielectric grains.
\newblock \emph{Astrophys. J.}, 186:\penalty0 705--714, 1973.
\newblock \doi{10.1086/152538}.

\bibitem[Spesyvtseva and Dholakia(2016)]{RN87}
Susan E.~Skelton Spesyvtseva and Kishan Dholakia.
\newblock Trapping in a material world.
\newblock \emph{ACS Photonics}, 3\penalty0 (5):\penalty0 719--736, 2016.
\newblock \doi{10.1021/acsphotonics.6b00023}.
\newblock URL \url{https://doi.org/10.1021/acsphotonics.6b00023}.

\bibitem[Vigoureux and Courjon(1992)]{RN75}
J.~M. Vigoureux and D.~Courjon.
\newblock Detection of nonradiative fields in light of the heisenberg
  uncertainty principle and the rayleigh criterion.
\newblock \emph{Appl. Opt.}, 31\penalty0 (16):\penalty0 3170--3177, Jun 1992.
\newblock \doi{10.1364/AO.31.003170}.
\newblock URL \url{http://ao.osa.org/abstract.cfm?URI=ao-31-16-3170}.

\bibitem[Novotny and Hecht(2012)]{RN57}
Lukas Novotny and Bert Hecht.
\newblock \emph{Principles of Nano-Optics}.
\newblock Cambridge University Press, Cambridge, 2012.
\newblock ISBN 9781107005464 1107005469.

\bibitem[Kotsifaki and {Nic Chormaic}(2019)]{KNC19}
D~Kotsifaki and S.~{Nic Chormaic}.
\newblock Plasmonic optical tweezers based on nanostructures: fundamentals,
  advances and prospect.
\newblock \emph{Nanophotonics}, 8\penalty0 (7):\penalty0 1227--1245, 2019.
\newblock \doi{10.1515/nanoph-2019-0151}.

\bibitem[Kawata and Sugiura(1992)]{RN76}
Satoshi Kawata and Tadao Sugiura.
\newblock Movement of micrometer-sized particles in the evanescent field of a
  laser beam.
\newblock \emph{Opt. Lett.}, 17\penalty0 (11):\penalty0 772--774, Jun 1992.
\newblock \doi{10.1364/OL.17.000772}.
\newblock URL \url{http://ol.osa.org/abstract.cfm?URI=ol-17-11-772}.

\bibitem[Okamoto and Kawata(1999)]{RN77}
K.~Okamoto and S.~Kawata.
\newblock Radiation force exerted on subwavelength particles near a
  nanoaperture.
\newblock \emph{Phys. Rev. Lett.}, 83:\penalty0 4534--4537, Nov 1999.
\newblock \doi{10.1103/PhysRevLett.83.4534}.
\newblock URL \url{https://link.aps.org/doi/10.1103/PhysRevLett.83.4534}.

\bibitem[Quidant et~al.(2005)Quidant, Petrov, and Badenes]{RN78}
Romain Quidant, Dmitri Petrov, and Gon\c{c}al Badenes.
\newblock Radiation forces on a rayleigh dielectric sphere in a patterned
  optical near field.
\newblock \emph{Opt. Lett.}, 30\penalty0 (9):\penalty0 1009--1011, May 2005.
\newblock \doi{10.1364/OL.30.001009}.
\newblock URL \url{http://ol.osa.org/abstract.cfm?URI=ol-30-9-1009}.

\bibitem[Volpe et~al.(2006)Volpe, Quidant, Badenes, and Petrov]{RN21}
G.~Volpe, R.~Quidant, G.~Badenes, and D.~Petrov.
\newblock Surface plasmon radiation forces.
\newblock \emph{Phys. Rev. Lett.}, 96\penalty0 (23):\penalty0 238101, 2006.
\newblock ISSN 0031-9007 (Print) 0031-9007 (Linking).
\newblock \doi{10.1103/PhysRevLett.96.238101}.
\newblock URL \url{https://www.ncbi.nlm.nih.gov/pubmed/16803408}.

\bibitem[A.~Maier(2007)]{Maier}
Stefan A.~Maier.
\newblock \emph{Plasmonics: Fundamentals and Applications}.
\newblock Springer, New York, 01 2007.
\newblock ISBN 978-0-387-33150-8.
\newblock \doi{10.1007/0-387-37825-1}.

\bibitem[Grigorenko et~al.(2008)Grigorenko, Roberts, Dickinson, and
  Zhang]{RN81}
A.~N. Grigorenko, N.~W. Roberts, M.~R. Dickinson, and Y.~Zhang.
\newblock Nanometric optical tweezers based on nanostructured substrates.
\newblock \emph{Nat. Photonics}, 2:\penalty0 365, Nov 2008.
\newblock \doi{10.1038/nphoton.2008.78}.
\newblock URL
  \url{https://www.nature.com/articles/nphoton.2008.78#supplementary-information}.

\bibitem[Righini et~al.(2009)Righini, Ghenuche, Cherukulappurath,
  Myroshnychenko, García~de Abajo, and Quidant]{RN82}
M.~Righini, P.~Ghenuche, S.~Cherukulappurath, V.~Myroshnychenko, F.~J.
  García~de Abajo, and R.~Quidant.
\newblock Nano-optical trapping of rayleigh particles and escherichia coli
  bacteria with resonant optical antennas.
\newblock \emph{Nano Lett.}, 9\penalty0 (10):\penalty0 3387--3391, 2009.
\newblock \doi{10.1021/nl803677x}.
\newblock URL \url{https://doi.org/10.1021/nl803677x}.
\newblock PMID: 19159322.

\bibitem[Chen et~al.(2012)Chen, Juan, Li, Maes, Borghs, Van~Dorpe, and
  Quidant]{RN15}
C.~Chen, M.~L. Juan, Y.~Li, G.~Maes, G.~Borghs, P.~Van~Dorpe, and R.~Quidant.
\newblock Enhanced optical trapping and arrangement of nano-objects in a
  plasmonic nanocavity.
\newblock \emph{Nano Lett.}, 12\penalty0 (1):\penalty0 125--32, 2012.
\newblock ISSN 1530-6992 (Electronic) 1530-6984 (Linking).
\newblock \doi{10.1021/nl2031458}.
\newblock URL \url{https://www.ncbi.nlm.nih.gov/pubmed/22136462}.

\bibitem[Pang and Gordon(2012)]{RN83}
Yuanjie Pang and Reuven Gordon.
\newblock Optical trapping of a single protein.
\newblock \emph{Nano Lett.}, 12\penalty0 (1):\penalty0 402--406, 2012.
\newblock \doi{10.1021/nl203719v}.
\newblock URL \url{https://doi.org/10.1021/nl203719v}.
\newblock PMID: 22171921.

\bibitem[Han et~al.(2018)Han, Truong, Thomas, and {Nic Chormaic}]{RN84}
Xue Han, Viet~Giang Truong, Prince~Sunil Thomas, and S\'{i}le {Nic Chormaic}.
\newblock Sequential trapping of single nanoparticles using a gold plasmonic
  nanohole array.
\newblock \emph{Photon. Res.}, 6\penalty0 (10):\penalty0 981--986, Oct 2018.
\newblock \doi{10.1364/PRJ.6.000981}.
\newblock URL
  \url{http://www.osapublishing.org/prj/abstract.cfm?URI=prj-6-10-981}.

\bibitem[Bethe(1944)]{RN5}
H.~A. Bethe.
\newblock Theory of diffraction by small holes.
\newblock \emph{Phys. Rev.}, 66\penalty0 (7-8):\penalty0 163--182, 1944.
\newblock ISSN 0031-899X.
\newblock \doi{10.1103/PhysRev.66.163}.

\bibitem[Ebbesen et~al.(1998)Ebbesen, Lezec, Ghaemi, Thio, and Wolff]{RN88}
T.~W. Ebbesen, H.~J. Lezec, H.~F. Ghaemi, T.~Thio, and P.~A. Wolff.
\newblock Extraordinary optical transmission through sub-wavelength hole
  arrays.
\newblock \emph{Nature}, 391:\penalty0 667, 1998.
\newblock \doi{10.1038/35570}.
\newblock URL \url{https://doi.org/10.1038/35570}.

\bibitem[Mart\'{\i}n-Moreno et~al.(2001)Mart\'{\i}n-Moreno, Garc\'{\i}a-Vidal,
  Lezec, Pellerin, Thio, Pendry, and Ebbesen]{RN90}
L.~Mart\'{\i}n-Moreno, F.~J. Garc\'{\i}a-Vidal, H.~J. Lezec, K.~M. Pellerin,
  T.~Thio, J.~B. Pendry, and T.~W. Ebbesen.
\newblock Theory of extraordinary optical transmission through subwavelength
  hole arrays.
\newblock \emph{Phys. Rev. Lett.}, 86:\penalty0 1114--1117, Feb 2001.
\newblock \doi{10.1103/PhysRevLett.86.1114}.
\newblock URL \url{https://link.aps.org/doi/10.1103/PhysRevLett.86.1114}.

\bibitem[de~Abajo(2002)]{RN85}
F.~J.~Garc\'{i}a de~Abajo.
\newblock Light transmission through a single cylindrical hole in a metallic
  film.
\newblock \emph{Opt. Express}, 10\penalty0 (25):\penalty0 1475--1484, Dec 2002.
\newblock \doi{10.1364/OE.10.001475}.
\newblock URL
  \url{http://www.opticsexpress.org/abstract.cfm?URI=oe-10-25-1475}.

\bibitem[Genet and Ebbesen(2007)]{RN6}
C.~Genet and T.~W. Ebbesen.
\newblock Light in tiny holes.
\newblock \emph{Nature}, 445\penalty0 (7123):\penalty0 39--46, 2007.
\newblock ISSN 1476-4687 (Electronic) 0028-0836 (Linking).
\newblock \doi{10.1038/nature05350}.
\newblock URL \url{https://www.ncbi.nlm.nih.gov/pubmed/17203054}.

\bibitem[Mestres et~al.(2016)Mestres, Berthelot, Acimovic, and Quidant]{RN22}
P.~Mestres, J.~Berthelot, S.~S. Acimovic, and R.~Quidant.
\newblock Unraveling the optomechanical nature of plasmonic trapping.
\newblock \emph{Light. Sci. Appl.}, 5\penalty0 (7):\penalty0 e16092, 2016.
\newblock ISSN 2047-7538 (Electronic) 2047-7538 (Linking).
\newblock \doi{10.1038/lsa.2016.92}.
\newblock URL \url{https://www.ncbi.nlm.nih.gov/pubmed/30167173}.

\bibitem[Sainidou and Garc\'{\i}a~de Abajo(2008)]{RN91}
R.~Sainidou and F.~J. Garc\'{\i}a~de Abajo.
\newblock Optically tunable surfaces with trapped particles in microcavities.
\newblock \emph{Phys. Rev. Lett.}, 101:\penalty0 136802, Sep 2008.
\newblock \doi{10.1103/PhysRevLett.101.136802}.
\newblock URL \url{https://link.aps.org/doi/10.1103/PhysRevLett.101.136802}.

\bibitem[Juan et~al.(2009)Juan, Gordon, Pang, Eftekhari, and Quidant]{RN7}
Mathieu~L. Juan, Reuven Gordon, Yuanjie Pang, Fatima Eftekhari, and Romain
  Quidant.
\newblock Self-induced back-action optical trapping of dielectric
  nanoparticles.
\newblock \emph{Nat. Phys.}, 5\penalty0 (12):\penalty0 915--919, 2009.
\newblock ISSN 1745-2473 1745-2481.
\newblock \doi{10.1038/nphys1422}.

\bibitem[Neumeier et~al.(2015)Neumeier, Quidant, and Chang]{RN11}
Lukas Neumeier, Romain Quidant, and Darrick~E. Chang.
\newblock Self-induced back-action optical trapping in nanophotonic systems.
\newblock \emph{New J. Phys.}, 17\penalty0 (12):\penalty0 123008, 2015.
\newblock ISSN 1367-2630.
\newblock \doi{10.1088/1367-2630/17/12/123008}.

\bibitem[Svoboda and Block(1994)]{RN121}
Karel Svoboda and Steven~M. Block.
\newblock Biological applications of optical forces.
\newblock \emph{Biophys. Biomol. Struct.}, 23:\penalty0 247--285, 1994.

\bibitem[Bui et~al.(2017)Bui, Stilgoe, Lenton, Gibson, Kashchuk, Zhang,
  Rubinsztein-Dunlop, and Nieminen]{RN122}
Ann A.~M. Bui, Alexander~B. Stilgoe, Isaac C.~D. Lenton, Lachlan~J. Gibson,
  Anatolii~V. Kashchuk, Shu Zhang, Halina Rubinsztein-Dunlop, and Timo~A.
  Nieminen.
\newblock Theory and practice of simulation of optical tweezers.
\newblock \emph{Journal of Quantitative Spectroscopy and Radiative Transfer},
  195:\penalty0 66--75, 2017.
\newblock ISSN 00224073.
\newblock \doi{10.1016/j.jqsrt.2016.12.026}.

\bibitem[Descharmes et~al.(2013)Descharmes, Dharanipathy, Diao, Tonin, and
  Houdre]{RN12}
N.~Descharmes, U.~P. Dharanipathy, Z.~Diao, M.~Tonin, and R.~Houdre.
\newblock Observation of backaction and self-induced trapping in a planar
  hollow photonic crystal cavity.
\newblock \emph{Phys. Rev. Lett.}, 110\penalty0 (12):\penalty0 123601, 2013.
\newblock ISSN 1079-7114 (Electronic) 0031-9007 (Linking).
\newblock \doi{10.1103/PhysRevLett.110.123601}.
\newblock URL \url{https://www.ncbi.nlm.nih.gov/pubmed/25166804}.

\bibitem[Bowman and Padgett(2013)]{RN123}
R.~W. Bowman and M.~J. Padgett.
\newblock Optical trapping and binding.
\newblock \emph{Rep Prog Phys}, 76\penalty0 (2):\penalty0 026401, 2013.
\newblock ISSN 1361-6633 (Electronic) 0034-4885 (Linking).
\newblock \doi{10.1088/0034-4885/76/2/026401}.
\newblock URL \url{https://www.ncbi.nlm.nih.gov/pubmed/23302540}.

\bibitem[Stevenson et~al.(2010)Stevenson, Gunn-Moore, and Dholakia]{RN124}
D.~J. Stevenson, F.~Gunn-Moore, and K.~Dholakia.
\newblock Light forces the pace: optical manipulation for biophotonics.
\newblock \emph{J Biomed Opt}, 15\penalty0 (4):\penalty0 041503, 2010.
\newblock ISSN 1560-2281 (Electronic) 1083-3668 (Linking).
\newblock \doi{10.1117/1.3475958}.
\newblock URL \url{https://www.ncbi.nlm.nih.gov/pubmed/20799781}.

\bibitem[Capitanio and Pavone(2013)]{RN125}
M.~Capitanio and F.~S. Pavone.
\newblock Interrogating biology with force: single molecule high-resolution
  measurements with optical tweezers.
\newblock \emph{Biophys J}, 105\penalty0 (6):\penalty0 1293--303, 2013.
\newblock ISSN 1542-0086 (Electronic) 0006-3495 (Linking).
\newblock \doi{10.1016/j.bpj.2013.08.007}.
\newblock URL \url{https://www.ncbi.nlm.nih.gov/pubmed/24047980}.

\bibitem[Nussenzveig(2018)]{RN126}
H.~M. Nussenzveig.
\newblock Cell membrane biophysics with optical tweezers.
\newblock \emph{Eur Biophys J}, 47\penalty0 (5):\penalty0 499--514, 2018.
\newblock ISSN 1432-1017 (Electronic) 0175-7571 (Linking).
\newblock \doi{10.1007/s00249-017-1268-9}.
\newblock URL \url{https://www.ncbi.nlm.nih.gov/pubmed/29164289}.

\bibitem[Sugiyama et~al.(2012)Sugiyama, Yuyama, and Masuhara]{RN127}
Teruki Sugiyama, Ken-Ichi Yuyama, and Hiroshi Masuhara.
\newblock Laser trapping chemistry : From polymer assembly to amino acid
  crystallization.
\newblock \emph{Acc. of Chem. Res.}, 45\penalty0 (11):\penalty0 1946--1954,
  2012.

\bibitem[Snook et~al.(2009)Snook, Harvey, Correia~Faria, and Gardner]{RN128}
R.~D. Snook, T.~J. Harvey, E.~Correia~Faria, and P.~Gardner.
\newblock Raman tweezers and their application to the study of singly trapped
  eukaryotic cells.
\newblock \emph{Integr Biol (Camb)}, 1\penalty0 (1):\penalty0 43--52, 2009.
\newblock ISSN 1757-9708 (Electronic) 1757-9694 (Linking).
\newblock \doi{10.1039/b815253e}.
\newblock URL \url{https://www.ncbi.nlm.nih.gov/pubmed/20023790}.

\bibitem[Chan(2013)]{RN129}
J.~W. Chan.
\newblock Recent advances in laser tweezers raman spectroscopy (ltrs) for
  label-free analysis of single cells.
\newblock \emph{J Biophotonics}, 6\penalty0 (1):\penalty0 36--48, 2013.
\newblock ISSN 1864-0648 (Electronic) 1864-063X (Linking).
\newblock \doi{10.1002/jbio.201200143}.
\newblock URL \url{https://www.ncbi.nlm.nih.gov/pubmed/23175434}.

\bibitem[Gillibert et~al.(2019)Gillibert, Balakrishnan, Deshoules, Tardivel,
  Magazzù, Donato, Maragò, Lamy~de La~Chapelle, Colas, Lagarde, and
  Gucciardi]{RN130}
Raymond Gillibert, Gireeshkumar Balakrishnan, Quentin Deshoules, Morgan
  Tardivel, Alessandro Magazzù, Maria~Grazia Donato, Onofrio~M. Maragò, Marc
  Lamy~de La~Chapelle, Florent Colas, Fabienne Lagarde, and Pietro~G.
  Gucciardi.
\newblock Raman tweezers for small microplastics and nanoplastics
  identification in seawater.
\newblock \emph{Environmental Science \& Technology}, 53\penalty0
  (15):\penalty0 9003--9013, 2019.
\newblock \doi{10.1021/acs.est.9b03105}.

\end{thebibliography}

\end{document}